\documentclass[aps,pra,reprint,superscriptaddress,showpacs,twocolumn]{revtex4}
\usepackage{amsfonts}
\usepackage{amsmath}
\usepackage{amssymb}
\usepackage{graphicx}
\usepackage{color}

\begin{document}

\title{Impact of anisotropy on the noncritical squeezing properties of \\
two-transverse-mode optical parametric oscillators}
\author{Carlos Navarrete-Benlloch}
\email{carlos.navarrete@mpq.mpg.de}
\affiliation{Max-Planck-Institut f\"ur Quantenoptik, Hans-Kopfermann-strasse 1, 85748
Garching, Germany}
\author{Germ\'an J. de Valc\'arcel}
\affiliation{Departament d'\`Optica, Universitat de Val\`encia, Dr. Moliner 50, 46100
Burjassot, Spain}
\date{\today}

\begin{abstract}
In a series of articles we studied the quantum properties of a degenerate
optical parametric oscillator tuned to the first family of transverse modes
at the subharmonic. We found that, for a cavity having rotational symmetry
respect to the optical axis, a TEM$_{10}$ mode with an arbitrary orientation
in the transverse plane is emitted above threshold. We proved then that
quantum noise induces a random rotation of this bright TEM$_{10}$ mode in the
transverse plane, while the mode orthogonal to it, the so-called dark mode, has perfect quadrature
squeezing irrespective of the distance to threshold (noncritical squeezing). This result was linked
to the spontaneous rotational symmetry breaking which occurs when the bright mode is generated, and here we analyze how the squeezing of
the dark mode is degraded when the cavity is not perfectly symmetric. We will show that large levels of squeezing
are still attainable for reasonable values of this anisotropy, with the advantage that the orientation of the bright and dark modes is basically fixed, a very attractive situation from the experimental point of view.
\end{abstract}

\maketitle

\textit{Introduction and previous results}. Squeezed states are a
fundamental resource both for high-precision measurements \cite%
{Goda08,Vahlbruch05,Treps02,Treps03} and quantum information protocols based
on continuous variables \cite{Braunstein05,Weedbrook12}. Nowadays, the
highest quality squeezed states are generated by using optical parametric
oscillators (OPOs) \cite{Mehmet10,Vahlbruch08,Takeno07}, which in essence
consist in a second order nonlinear crystal embedded in an optical cavity.
When pumped with a field of frequency $2\omega_{0}$, photons of frequency $%
\omega_{0}$ are generated in the crystal via the process of parametric
down-conversion, and large levels of squeezing are found in this
down-converted field only when the OPO is operated close to threshold \cite%
{Meystre91}.

In recent works we have analyzed the properties of such devices when several
down-conversion channels corresponding to different temporal \cite%
{Valcarcel06,Patera10,Patera12}, spatial \cite%
{Patera10,Perez06,Perez07,Navarrete08,Navarrete09,Navarrete10,Navarrete11},
or polarization \cite{Garcia10} modes, are available for a given pump mode,
obtaining a so-called multi-mode OPO. We have shown that the properties of
these devices can be understood in terms of two fundamental phenomena \cite%
{NavarretePhDthesis}: pump clamping \cite{Navarrete09} and spontaneous
symmetry breaking \cite{Perez06,Perez07,Navarrete08,Navarrete10,Garcia10}.

In a series of papers \cite{Navarrete08,Navarrete10,Navarrete11}, we have
studied in detail the quantum properties derived from the spontaneous
symmetry breaking process by means of a particular example: the
two-transverse-mode degenerate OPO. The particularity of this OPO is that,
while a TEM$_{00}$ mode is resonant at frequency $2\omega_{0}$ as usual, its
cavity is tuned to the first family of transverse modes at the subharmonic
frequency $\omega_{0}$, so that the pump photons can be down-converted into
pairs of either TEM$_{10}$ or TEM$_{01}$ photons. This cavity configuration
has been already tested in several experiments \cite%
{Lassen09,Janousek09,Chalopin10}, although none of them have enter the
regime in which the first family of transverse modes oscillates above
threshold, what could be accomplished by using the multi-Gaussian pump
technique introduced in \cite{Patera12}.

When this OPO is pumped at frequency $2\omega_{0}$ above some threshold, we
proved that classical emission takes place in a TEM$_{10}$ mode with an
arbitrary orientation in the transverse plane, this indeterminacy coming
from the invariance of the system under rotations around the optical axis 
\cite{Navarrete08,Navarrete10}. One can distinguish then between a \textit{%
bright} and a \textit{dark\ }mode, which are, respectively, the TEM$_{10}$
mode in which mean field emission occurs and its orthogonal.

As for the quantum phenomena, we showed two fundamental properties \cite%
{Navarrete08,Navarrete10}. First, as any orientation of the bright mode is
allowed, quantum noise is able to act freely on this variable; in
particular, we found that the variance of this orientation increases
linearly with time, so that, eventually, the orientation of the mode becomes
completely uncertain: one says that quantum noise tends to restore the
symmetry broken at the mean field level. On the other hand, while the
properties of the bright mode are the same as those of the single-mode DOPO
(it shows large squeezing levels only when working close to threshold), the
dark mode was shown to have perfect quadrature squeezing at any pumping
level above threshold, a result which we interpreted in terms of an
angle-angular momentum uncertainty relation \cite{Navarrete08,Navarrete10}.

However, since the orientation of the dark mode is uncertain (in the quantum
mechanical sense), it seems highly unlikely to match the local oscillator to
it in a homodyne detection experiment \cite{Navarrete10}, and hence, we
studied in \cite{Navarrete11} the possibility of locking its orientation by
seeding the cavity with a TEM$_{10}$ coherent beam at the subharmonic, what
constitutes an explicit breaking of the system's symmetry, and hence
degrades the level of squeezing found in the dark mode (which corresponds to
the TEM$_{01}$ mode when locking is accomplished). We showed that large
levels of squeezing can still be obtained for reasonable values of the
injection's power.

In this brief report we study how the quantum properties of this system are
affected by two sources of anisotropy: a tilt of the nonlinear crystal (or,
equivalently, the introduction of a tilted dielectric slab in the cavity),
and a possible astigmatism of the mirrors. The first one is particularly
interesting, since it allows the locking of the bright mode's orientation in
a very simple, noninvasive, and controllable way. In addition, a tilting of
the nonlinear crystal is sometimes desirable for phase-matching purposes 
\cite{Boyd}, and our study will reveal the limits that the degradation of
the quantum properties impose on this technique.

\textit{Describing transverse anisotropy}. The fact that both the TEM$_{10}$
and TEM$_{01}$ modes (or any TEM$_{10}$ mode\ with an arbitrary orientation
in the transverse plane) have the same resonant frequency comes from the
system's invariance with respect to rotations around the optical axis (taken
as the $z$ axis). As we show below, the main effect of any source of
transverse anisotropy is the introduction of a detuning between the TEM$%
_{10} $ and TEM$_{01}$ modes. In what follows, we will take $\omega_{0}$ to
be the frequency of the TEM$_{10}$ mode, while the TEM$_{01}$ mode will
resonate at frequency $\omega_{0}+\Delta$, being $\Delta$ the detuning
introduced by the anisotropy. As long as $\Delta/\omega_{0}$ remains small,
any other effect such as the change in the spatial form of the modes turns
out to be irrelevant for our purposes.

We will consider two different sources of anisotropy. The first one consists
in allowing one of the mirrors to have some astigmatism; we model this by
introducing different curvature radii $R_{x}$ and $R_{y}$ in the $x$ and $y$
directions, respectively, so that the mirror becomes ellipsoidal. The second
one consists on allowing the $\chi^{\left( 2\right) }$ crystal (of length $l_\mathrm{c}$ and refractive index $n_\mathrm{c}$) to be
slightly tilted with respect to the cavity axis; we will assume that tilt
occurs in the $zx$ plane with angle $\beta$ with respect to the optical
axis, what introduces different effective lengths along the $x$ and $y$
directions given by \cite{Hanna69} 
\begin{subequations}
\begin{align}
L_{\mathrm{eff},x} & =L-l_{\mathrm{c}}\left[ \left\vert \cos\beta\right\vert
+\frac{\sin^{2}\beta\left( 2n_{\mathrm{c}}^{2}-\sin^{2}\beta\right) -n_{%
\mathrm{c}}^{2}}{\left( n_{\mathrm{c}}^{2}-\sin^{2}\beta\right) ^{3/2}}%
\right] , \\
L_{\mathrm{eff},y} & =L-l_{\mathrm{c}}\left[ \left\vert \cos\beta\right\vert
-\frac{\cos^{2}\beta}{\left( n_{\mathrm{c}}^{2}-\sin^{2}\beta\right) ^{1/2}}%
\right] \text{,}
\end{align}
$L$ being the cavity length. Note that instead of tilting the nonlinear crystal, one can obtain exactly
the same effect by introducing a tilted dielectric slab with anti-reflecting
coating.

Both types of anisotropies have the effect of changing the $g$-parameters \cite{gParameters} of
the cavity in a different way for the $x$ and $y$ transverse directions,
hence turning the cavity into a composition of two orthogonal 1D cavities
with $g$-parameters $g_{1x}=1-L_{\mathrm{eff},x}/R$ ($g_{2x}=1-L_{\mathrm{eff%
},x}/R_{x}$) and $g_{1y}=1-L_{\mathrm{eff},y}/R$ ($g_{2y}=1-L_{\mathrm{eff}%
,y}/R_{y}$) for mirror 1 (2). Note that we have taken mirror 2 as the
astigmatic one for definiteness. Under these conditions, the resonance
frequency of a TEM$_{mn}$ mode with longitudinal index $q$ is given by 
\end{subequations}
\begin{align}
\omega_{qmn} & =\frac{\pi c}{L_{\mathrm{opt}}}\left( q+\frac{m+1/2}{\pi }%
\arccos\sqrt{g_{1x}g_{2x}}\right.  \notag \\
& \left. +\frac{n+1/2}{\pi}\arccos\sqrt{g_{1y}g_{2y}}\right) ,
\label{AnisoResonances}
\end{align}
where $L_{\mathrm{opt}}=L+[ (n_{\mathrm{c}}^{2}-\sin^{2}\beta)^{1/2}-\left%
\vert\cos\beta\right\vert ] l_{\mathrm{c}}$, and where we have assumed that
the curvature radii of the mirrors are larger than the corresponding
effective lengths as usually happens in OPO experiments, so that all the $g$%
-parameters are larger than one. Hence the detuning between the TEM$_{10}$
and TEM$_{01}$ modes is given by%
\begin{equation}
\Delta=c\left( \arccos\sqrt{g_{1y}g_{2y}}-\arccos\sqrt{g_{1x}g_{2x}}%
\right)/L_{\mathrm{opt}} .  \label{Detuning}
\end{equation}

\textit{The model equations}. As usual, the cavity is pumped by an external
laser resonant at frequency $2\omega_{0}$ with a TEM$_{00}$ mode of the
cavity; pump photons are down-converted in the $\chi^{\left( 2\right) }$
crystal into pairs of photons laying either in the TEM$_{10}$ or TEM$_{01}$
mode. The (interaction picture) Hamiltonian is written as%
\begin{equation}
\hat{H}=\hbar\Delta\hat{a}_{y}^{\dagger}\hat{a}_{y}+i\hbar \mathcal{E}_{%
\mathrm{p}}\hat{a}_{0}^{\dagger}+i\hbar\chi\hat{a}_{0}(\hat{a}%
_{x}^{\dagger2}+\hat{a}_{y}^{\dagger2})/2+\mathrm{H.c.},
\end{equation}
being $\hat{a}_{0}$, $\hat{a}_{x}$, and $\hat{a}_{y}$ the annihilation
operators for TEM$_{00}$, TEM$_{10}$, and TEM$_{01}$ photons, respectively. $%
\mathcal{E}_{\mathrm{p}}$ is proportional to the amplitude of the pumping
laser, and we take it as real without loss of generality. The nonlinear
coupling $\chi$ is proportional to the nonlinear susceptibility of the $%
\chi^{\left( 2\right) }$ crystal and the overlapping transverse integral
between the three modes involved in the particular down-conversion process,
and is the same for the TEM$_{10}$ and TEM$_{01}$ modes except for
corrections of order $\Delta/\omega_{0}$, which can be neglected for optical
frequencies and reasonable anisotropies.

\begin{figure*}[t]
\includegraphics[width=\textwidth]{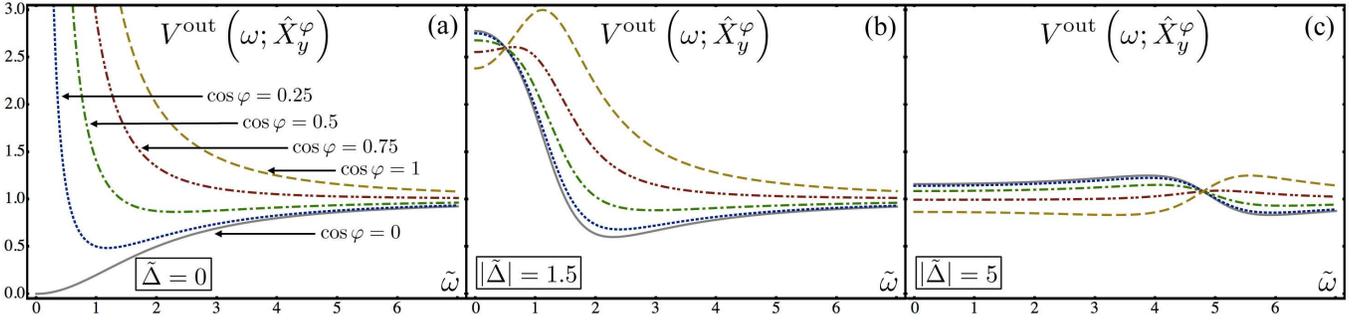}
\caption{Noise spectrum of the TEM$_{01}$ quadratures indicated in the plot
as a function of the noise frequency. Note that the minimum noise is always
achieved in the $\protect\varphi =\protect\pi /2$ quadrature; interestingly,
it is easy to prove from (\protect\ref{NoiseSpectrumTEM01}) that for $\tilde{%
\Delta}^{2}>2$ this quadrature has more noise than the $\protect\varphi =0$
quadrature at $\protect\omega =0$ (b,c). Indeed, the latter can even show
squeezing for $\tilde{\Delta}^{2}>4$; in particular, for $4<\tilde{\Delta}%
^{2}<2(3+\protect\sqrt{5})$ its optimum squeezing is found at $\protect%
\omega =0$, while for $\tilde{\Delta}^{2}>2(3+\protect\sqrt{5})$ it is found
at $\protect\omega ^{2}=\tilde{\Delta}^{2}-2|\tilde{\Delta}|-4$, and has the
same value as the optimum squeezing in the $\protect\varphi =\protect\pi /2$
quadrature (c).}
\label{Fig1}
\end{figure*}

As usual, we incorporate losses through the partially transmitting mirror by
introducing the decay rates $\gamma_{\mathrm{p}}$ and $\gamma_{\mathrm{s}}$
in the master equation for the pump and signal modes, respectively. Then we
map the master equation into a set of stochastic Langevin equations within
the positive \textit{P} representation; given the resemblance of our
Hamiltonian to that of \cite{Navarrete11}, we can just take the Langevin
equations of that reference but including a detuning $\Delta$ for the TEM$%
_{01}$ mode and no subharmonic injection, 
\begin{subequations}
\label{Langevin}
\begin{align}
\dot{\alpha}_{0} & =\mathcal{E}_{\mathrm{p}}-\gamma_{\mathrm{p}}\alpha
_{0}-\chi\left( \alpha_{x}^{2}+\alpha_{y}^{2}\right) /2, \\
\dot{\alpha}_{x} & =-\gamma_{\mathrm{s}}\alpha_{x}+\chi\alpha_{0}\alpha
_{x}^{+}+\sqrt{\chi\alpha_{0}}\eta_{x}\left( t\right) , \\
\dot{\alpha}_{y} & =-\left( \gamma_{\mathrm{s}}+i\Delta\right) \alpha
_{y}+\chi\alpha_{0}\alpha_{y}^{+}+\sqrt{\chi\alpha_{0}}\eta_{y}\left(
t\right) ,
\end{align}
plus the equations for $\alpha_{m}^{+}$ ($m=0,x,y$), which are like those
for $\alpha_{m}$ with the changes $\alpha_{m}\longleftrightarrow%
\alpha_{m}^{+}$, $\eta_{m}\left( t\right) \rightarrow\eta_{m}^{+}\left(
t\right) $ and $i\rightarrow-i$. $\eta_{m}\left( t\right) $ and $\eta
_{m}^{+}\left( t\right) $ are real, Gaussian noises with zero mean and
non-zero correlations 
\end{subequations}
\begin{equation}
\left\langle \eta_{m}\left( t\right) \eta_{m^{\prime}}\left( t^{\prime
}\right) \right\rangle =\left\langle \eta_{m}^{+}\left( t\right)
\eta_{m^{\prime}}^{+}\left( t^{\prime}\right) \right\rangle =\delta
_{m,m^{\prime}}\delta\left( t-t^{\prime}\right) .  \label{ruido}
\end{equation}

These equations allow us to obtain the noise spectrum of any quadrature $%
\hat{X}_{m}^{\varphi}=\exp\left( -i\varphi\right) \hat{a}_{m}+\exp\left(
i\varphi\right) \hat{a}_{m}^{\dagger}$ as%
\begin{equation}
V^{\mathrm{out}}\left( \omega;\hat{X}_{m}^{\varphi}\right)
=1+S_{m}^{\varphi}\left( \omega\right) ,
\end{equation}
with the squeezing spectrum given by%
\begin{equation}
S_{m}^{\varphi}\left( \omega\right) =2\gamma_{m}\int_{-\infty}^{+\infty
}d\tau e^{-i\omega\tau}\left\langle \delta X_{m}^{\varphi}\left( 0\right)
\delta X_{m}^{\varphi}\left( \tau\right) \right\rangle .
\end{equation}
In this expression $\delta X=X-\left\langle X\right\rangle $, and $%
X_{m}^{\varphi}=\exp\left( -i\varphi\right) \alpha_{m}+\exp\left(
i\varphi\right) \alpha_{m}^{+}$ is the stochastic counterpart of the
corresponding quadrature operator. For a vacuum or a coherent state $%
S_{m}^{\varphi}\left( \omega\right) =0$, and hence $V^{\mathrm{out}}\left( 
\bar{\omega};X_{m}^{\varphi}\right) <1$ signals squeezing in the
corresponding quadrature at noise frequency $\bar{\omega}$.

\textit{The classical picture}. The classical dynamic equations of the
system are recovered from (\ref{Langevin}) by removing the noise terms and
replacing $\alpha_{m}^{+}$ by $\alpha_{m}^{\ast}$. The resulting equations
have two types of stationary solutions as is well known in OPOs: for $%
\mathcal{E}_{\mathrm{p}}$ below some threshold value the down-converted
modes are switched off, while above that threshold one of them is switched
on. Which mode wins the nonlinear competition is dictated solely by which
one has the lowest threshold \cite{Navarrete09}. In the current case the TEM$%
_{10}$ mode is on resonance and its threshold corresponds to that of the
single-mode DOPO, that is, $\mathcal{E}_{\mathrm{th,}x}=\gamma_{\mathrm{p}%
}\gamma_{\mathrm{s}}/\chi$; on the other hand, the TEM$_{01}$ mode is
detuned and hence its threshold is given by $\mathcal{E}_{\mathrm{th,}y}=$ $%
\sqrt{1+\Delta ^{2}/\gamma_{\mathrm{s}}^{2}}\mathcal{E}_{\mathrm{th,}x}$ 
\cite{Savage87,Fabre90}. Hence, the TEM$_{10}$ mode has the lowest
threshold, and thus above threshold the only stable, stationary solution is
given by 
\begin{subequations}
\label{ClassicalSol}
\begin{align}
\bar{\alpha}_{0} & =\gamma_{\mathrm{s}}/\chi\text{, }\bar{\alpha}_{y}=0 \\
\bar{\alpha}_{x} & =\pm\rho\text{ with }\rho=\sqrt{2\left(\mathcal{E}_{%
\mathrm{p}}-\mathcal{E}_{\mathrm{th}}\right)/\chi },
\end{align}
showing that the anisotropy fixes the bright and dark TEM modes to the $x$
and $y$ axis, respectively, what is very convenient from the detection point
of view.

\textit{Squeezing properties of the dark mode}. One of the goals of this
work is to show that the quantum properties of the dark mode, which has
perfect squeezing irrespective of the distance to threshold in an isotropic
cavity, remain useful for reasonable values of the anisotropy. Hence, we
study now the squeezing properties of the TEM$_{01}$ mode, which is the dark
mode in the current setup.

As usual, we linearize the Langevin equations (\ref{Langevin}) around the
classical steady state solution (\ref{ClassicalSol}); as the TEM$_{01}$ mode
is off at the classical level, this aims to change the pump variables $%
\left\{ \alpha_{0},\alpha_{0}^{+}\right\} $ by their corresponding mean
field values $\left\{ \bar{\alpha}_{0},\bar{\alpha}_{0}^{\ast}\right\} $\ in
the equations for the $\left\{ \alpha_{y},\alpha_{y}^{+}\right\} $ modes,
arriving to 
\end{subequations}
\begin{equation}
\boldsymbol{\dot{\alpha}}_{y}=\mathcal{L}\boldsymbol{\alpha}_{y}+\sqrt {%
\gamma_{\mathrm{s}}}\boldsymbol{\eta}_{y}\left( t\right) \text{,}
\end{equation}
with $\boldsymbol{\alpha}_{y}=\mathrm{col}\left( \alpha_{y},\alpha
_{y}^{+}\right) $, $\boldsymbol{\eta}_{y}=\mathrm{col}\left( \eta
_{y},\eta_{y}^{+}\right) $, and%
\begin{equation}
\mathcal{L}=%
\begin{pmatrix}
-\gamma_{\mathrm{s}}+i\Delta & \gamma_{\mathrm{s}} \\ 
\gamma_{\mathrm{s}} & -\gamma_{\mathrm{s}}+i\Delta%
\end{pmatrix}
.
\end{equation}

Following Collett and Walls \cite{Collet85}, the squeezing spectrum of any TEM%
$_{01}$ quadrature can then be obtained as%
\begin{equation}
S_{y}^{\varphi}\left( \omega\right) =2\gamma_{\mathrm{s}}\mathrm{Re}\left\{
e^{-2i\varphi}\mathcal{S}_{11}\left( \omega\right) +\mathcal{S}_{12}\left(
\omega\right) \right\} ,
\end{equation}
where the spectral covariance matrix is given in our case by $\mathcal{S}%
\left( \omega\right) =\gamma_{\mathrm{s}}\left( \mathcal{L}+i\omega\mathcal{I%
}\right) ^{-1}\left( \mathcal{L}^{T}-i\omega\mathcal{I}\right) ^{-1}$, being 
$\mathcal{I}$ the $2\times2$ identity matrix. A straightforward algebraic
manipulation allows us to write the noise spectrum as%
\begin{equation}
V^{\mathrm{out}}\left( \omega;\hat{X}_{y}^{\varphi}\right) =1+\frac{4\left( 
\tilde{\Delta}^{2}-\tilde{\omega}^{2}\right) +8\left( 2-\tilde{\Delta}^{2}+%
\tilde{\omega}^{2}\right) \cos^{2}\varphi}{4\tilde{\omega}^{2}+\left( \tilde{%
\Delta}^{2}-\tilde{\omega}^{2}\right) ^{2}},  \label{NoiseSpectrumTEM01}
\end{equation}
where $\tilde{\Delta}=\Delta/\gamma_{\mathrm{s}}$ and $\tilde{\omega}%
=\omega/\gamma_{\mathrm{s}}$. The spectrum is showed in Fig. 1 for different
values of the parameters. A simple inspection of this expression shows that $%
\varphi=\pi/2$ is the maximally squeezed quadrature for any combination of
the parameters; also, it is straightforward to show that for a given value
of the detuning, there exists an optimum detection frequency $\tilde{\omega }%
_{\mathrm{opt}}^{2}=\tilde{\Delta}^{2}+2|\tilde{\Delta}|$ for which
squeezing is maximum, $V_{\mathrm{opt}}^{\mathrm{out}}=|\tilde{\Delta}|/(1+|%
\tilde{\Delta}|)$ in particular. This expression shows that more than 90\%
of squeezing ($V_{\mathrm{opt}}^{\mathrm{out}}<0.1$) can be obtained as long
as $|\tilde{\Delta}|<0.1$.

In order to get an idea of the detuning induced by the different sources of
anisotropy considered in this brief report, in Fig. 2 we show the dependence
of $|\tilde{\Delta}|$ with the tilting angle $\beta$ of the crystal and the
ellipticity parameter $\varepsilon=1-R_{y}/R_{x}$ of the astigmatic mirror,
for typical OPO parameters. As expected, we observe that there exists a
limit to how much the crystal can be tilted if we want to satisfy the
condition $\tilde{\Delta}<0.1$ ($\sim$6 degrees in the example), what
imposes restrictions on techniques such as angular phase matching \cite{Boyd}%
; similarly, this condition imposes restrictions on the astigmatism of the
mirror (an ellipticity below $\sim$0.1\%, in the example), although this is
not a strong limitation since nowadays commercial mirrors can be made
virtually perfectly spherical.

In the limit of small $\beta$ and $\varepsilon$, that is, small anisotropy,
one can find the following simple expression for the detuning to the leading
order in these parameters:%
\begin{equation}
\tilde{\Delta}=\frac{2}{\mathcal{T}}\left[ \frac{2l_{\mathrm{c}}(n_{\mathrm{c%
}}^{2}-1)}{Rn_{\mathrm{c}}^{3}\sqrt{1-g^{2}}}\beta^{2}+\sqrt{\frac{1-g}{1+g}}%
|\varepsilon|\right] ,
\end{equation}
where we have taken $R_{x}=R$ to simplify the expression, $g$ is the $g$%
-parameter of the cavity in the absence of anisotropy, and we have used $%
\gamma_{\mathrm{s}}=c\mathcal{T}/4L_{\mathrm{opt}}$, being $\mathcal{T}$ the
transmissivity of the output-coupling mirror.

\textit{Connection to the orientation's uncertainty.} The second goal of
this report is to show how the indeterminacy of the bright mode's
orientation decreases as the anisotropy increases, an analysis aimed to
reinforce its relation with the squeezing of the dark mode \cite%
{Navarrete08,Navarrete10}.

Following previous works \cite{Navarrete08,Navarrete10}, we now take the
orientation of the bright mode as a dynamic variable $\theta (t)$, and
expand the stochastic amplitudes as 
\begin{subequations}
\label{AlphaThetaExpansion}
\begin{align}
\alpha _{x}(t)& =\rho \cos \theta (t)+b_{x}(t)\cos \theta (t)+b_{y}(t)\sin
\theta (t), \\
\alpha _{y}(t)& =\rho \sin \theta (t)+b_{x}(t)\sin \theta (t)+b_{y}(t)\cos
\theta (t),
\end{align}%
and similarly for the $\alpha _{m}^{+}$ amplitudes. The difference with
respect to our previous works is that, in addition to the fluctuations $b_{m}
$ and the orientation's time derivative $\dot{\theta}$, now we can take the
orientation $\theta $ itself as a small quantity, as it is classically
locked to $\theta =0$; we will come back to this point at the end of this
section. Linearization of (\ref{AlphaThetaExpansion}) with respect to these
variables leads to $\alpha _{x}\approx \rho +b_{x}$ and $\alpha _{y}=\rho
\theta +b_{y}$, and hence all the information about the orientation $\theta $
is contained in the dark mode's evolution equations, which can be written as 
\end{subequations}
\begin{equation}
\rho \mathbf{u}_{0}\dot{\theta}+\mathbf{\dot{b}}_{y}=-i\rho \Delta \mathbf{u}%
_{1}\theta -\mathcal{L}\mathbf{b}_{y}+\sqrt{\gamma _{\mathrm{s}}}\boldsymbol{%
\eta }_{y}\left( t\right) \text{,}
\end{equation}%
with $\mathbf{u}_{0}=\mathrm{col}\left( 1,1\right) $ and $\mathbf{u}_{1}=%
\mathrm{col}\left( 1,-1\right) $. Now, projecting this linear system onto $%
\mathbf{u}_{0}$ and $\mathbf{u}_{1}$, and calling $c_{j}=\mathbf{u}_{j}\cdot 
\mathbf{b}_{y}$ (we take $c_{0}=0$ as usual \cite{Navarrete08,Navarrete10}
to compensate for the excess of variables), we get the linear system%
\begin{equation}
\mathbf{\dot{x}}=-\mathcal{M}\mathbf{x}+\sqrt{2\gamma _{\mathrm{s}}}%
\boldsymbol{\eta }\left( t\right) ,
\end{equation}%
with%
\begin{equation}
\mathbf{x=}%
\begin{pmatrix}
2\rho \theta  \\ 
c_{1}%
\end{pmatrix}%
,\ \text{ and \ }\mathcal{M}=%
\begin{pmatrix}
0 & i\Delta  \\ 
i\Delta  & 2\gamma _{\mathrm{s}}%
\end{pmatrix}%
,
\end{equation}%
and where $\boldsymbol{\eta }\left( t\right) $ contains two real,
independent noises which satisfy the usual statistical properties (\ref%
{ruido}).

\begin{figure}[t]
\includegraphics[width=\columnwidth]{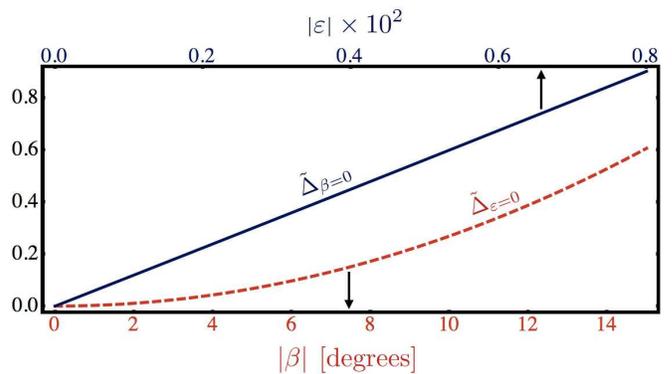}
\caption{Normalized detuning $\tilde{\Delta}$ as a function of the tilting
angle of the nonlinear crystal $\protect\beta $ and the ellipticity of the
astigmatic mirror $\protect\varepsilon $. The solid and dashed lines
correspond, respectively, to the case of a perfectly aligned crystal and a
perfectly spherical mirror. Parameters: $R=R_{x}=2L$, $l_{\mathrm{c}}=0.1L$, 
$n_{\mathrm{c}}=2$, and $\mathcal{T}=0.01$.}
\label{Fig2}
\end{figure}

This equation contains all the information about the evolution of the
orientation $\theta $, and it is in particular very simple, for example by
diagonalizing $\mathcal{M}$, to obtain the following long time term variance
from it%
\begin{equation}
V_{\theta }^{\infty }=\langle \theta ^{2}(t\rightarrow \infty )\rangle =%
\frac{1}{2\rho ^{2}\tilde{\Delta}^{2}},
\end{equation}%
which, as expected, shows that the quantum indetermination of the bright
mode's orientation monotonically decreases with the level of anisotropy.
Note that this expression has been found by assuming that $\theta $ does not
make large excursions from $\theta =0$, that is, it is valid as long as $%
V_{\theta }^{\infty }\ll 1$; note however that $\rho ^{2}$ approximately
gives the number of photons contained in the bright mode, and hence this
approximation is consistent even for small normalized detunings $\tilde{%
\Delta}$ as long as one works sufficiently above threshold.

\textit{Conclusions}. In this brief report we have analyzed the impact that
transverse anisotropy has onto the quantum properties of the
two-transverse-mode DOPO \cite{Navarrete08,Navarrete10}, showing that a
small amount of it can serve to lock the orientation of the bright and dark
modes (very beneficial for experiments) while still allowing for large
levels of squeezing in the dark mode.

\begin{acknowledgments}
We thank Eugenio Rold\'{a}n for fruitful discussions. This work has been
supported by the Spanish Government and the European Union FEDER through
Project FIS2011-26960. C.N.-B. acknowledges the financial support of the
Future and Emerging Technologies (FET) programme within the Seventh
Framework Programme for Research of the European Commission under the
FET-Open grant agreement MALICIA, number FP7-ICT-265522, and of the
Alexander von Humboldt Foundation through its Fellowship for Postdoctoral
Researchers.
\end{acknowledgments}

\end{document}